\documentclass{iopart}

\usepackage{iopams}
\expandafter\let\csname equation*\endcsname\relax
\expandafter\let\csname endequation*\endcsname\relax

\usepackage{amsmath,amsfonts,amssymb}

\usepackage[utf8]{inputenc} 
\usepackage[T1]{fontenc}    
\usepackage{xfrac}
\usepackage{hyperref}       
\usepackage{bm}
\usepackage{url}            
\usepackage{booktabs}       
\usepackage{amsfonts}       
\usepackage{nicefrac}       
\usepackage{microtype}      
\usepackage{lipsum}
\usepackage{graphicx}
\usepackage{dcolumn}
\usepackage{bm}
\usepackage[super]{cite}

\newcommand{\norm}[1]{\left\lVert#1\right\rVert}
\DeclareMathAlphabet{\mathpzc}{OT1}{pzc}{m}{it}

\newcommand{\beginsupplement}{%
        \setcounter{table}{0}
        \renewcommand{\thetable}{S\arabic{table}}%
        \setcounter{figure}{0}
        \renewcommand{\thefigure}{S\arabic{figure}}%
     }

\begin{document}
\title{On the role of gradients for machine learning of molecular energies and forces}


\author{
    Anders S. Christensen \& O. Anatole von Lilienfeld$^{*}$
    }
\address{
    Institute of Physical Chemistry\\
    National Center for Computational Design and \\
    Discovery of Novel Materials (MARVEL) \\
    Department of Chemistry \\
    University of Basel\\ 
    Klingelbergstrasse 80\\
    CH-4056 Basel, Switzerland\\
    $^{*}$\url{anatole.vonlilienfeld@unibas.ch}
}


\begin{abstract}
The accuracy of any machine learning potential can only be as good as the data used in the fitting process.
The most efficient model therefore selects the training data that will yield the highest accuracy compared to the cost of obtaining the training data.
We investigate the convergence of prediction errors of quantum machine learning models for organic molecules trained on energy and force labels, two common data types in molecular simulations.
When training and predicting on different geometries corresponding to the same single molecule, we find that the inclusion of atomic forces in the training data increases the accuracy of the predicted energies and forces 7-fold, compared to models trained on energy only.
Surprisingly, for models trained on sets of organic molecules of varying size and composition in non-equilibrium conformations, inclusion of forces in the training does not improve the predicted energies of unseen molecules in new conformations. 
Predicted forces, however, also improve about 7-fold.
For the systems studied, we find that force labels and energy labels contribute equally \textit{per} label to the convergence of the prediction errors. 
Choosing to include derivatives such as atomic forces in the training set or not should thus depend on, not only on the computational cost of acquiring the force labels for training, but also on the application domain, the property of interest, and the desirable size of the machine learning model. 
Based on our observations we describe key considerations for the creation of datasets for potential energy surfaces of molecules which maximize the efficiency of the resulting machine learning models.
%
%
%
%
%
%
%
%
%
\end{abstract}


\section{Introduction}

In recent years, machine learning models have become increasingly popular as methods to approximate potential energy surfaces of molecules.
These models range from classical learning methods such as kernel methods to methods based on deep neural networks.\cite{GAPtutorial,Neuralnetworks_Scheffler2004,Neuralnetworks_BehlerParrinello2007,BehlerPerspective2016,ANI_IsayevRoitberg2017,Botu2015,RampiMLQMMM,Botu2016,Huan2017,Zhenwei2015,gubaev2018machine,SNAP_Aidan2015,CovariantKernelsSandro2016,Glielmo2018,schutt2018schnet,schutt2019schnetpack,grisafi2018symmetry,DeepMDZhang2018,UnkePhysNet2019}
A common denominator for these data-driven models is that they require an adequate training set in order to yield predictions of sufficient accuracy.
It is thus clear that informed and rational selection of training data is paramount to proper optimization of the data-efficiency of the machine learning models.

For machine learning models describing potential energy surfaces, two types of data seem particularly convenient as training data: single-point energies and atomic force vectors.
However, it has not yet been fully demonstrated when and to which degree the inclusion of force labels in the training set truly leads to an improvement of the accuracy of the trained model versus energy labels. It is even possible to find somewhat conflicting information in literature.
For example, the GDML and sGDML methods achieve state-of-the-art accuracy in certain cases for the MD17 benchmark dataset, despite training only on force labels, ignoring the energy labels.\cite{chmiela2017machine,Chmiela2019sGDML,Christensen2020FCHL19}
In contrast, the  HIP-NN neural network---when only trained on energy labels---ostensibly achieves similar predictive accuracy to the DTNN, SchNet and PhysNet neural networks, when these are trained on both forces {\em and} energy labels for 50K molecules from the MD17 dataset, despite the fact that HIP-NN is trained using far fewer training labels.\cite{Lubber2018HIP,DTNN2017,schutt2018schnet,UnkePhysNet2019}
Similarly, the family of ANI datasets have resulted in successful potential energy models for general organic chemistry solely trained on single-point energies on geometries distorted along normal-modes, although the corresponding gradient information would not have been much more costly to obtain at the density-functional theory (DFT) level employed.\cite{smith2017ani,ANI_IsayevRoitberg2017,smith2018less,smith2020ani1cxx}

Consequently, it is not clear from literature which is the best strategy for designing a dataset for the most accurate machine learning potentials for chemical compounds within a given budget of computational resources.
In order to shed more light on this matter, here we investigate how the predictive accuracy of machine learning models is affected by inclusion of derivative information---namely atomic force vectors in addition to the energy---in the training set.
More specifically, we investigate the nine possible combinations resulting from the inclusion of functions, derivatives, and functions plus derivatives in loss-functions used for training,  as well as in error measures used for evaluating the prediction error. 

In order to \textit{learn}, the predictive accuracy of the trained machine \textit{must} strictly increase with an increasing amount of training data, aside from statistical variation.
This holds true as long as the training data is noiseless, and the machine learning model is flexible enough to be able to properly account for all variation in the data.\cite{vapnik2013book,QMLessayAnatole}
This has previously been demonstrated numerically with the leading term of the prediction error decreasing according to a power-law with number of training samples, $N$, for kernel ridge regression~\cite{vapnik1994learningcurves}, as well as for neural networks\cite{StatError_Muller1996}, i.e.:
\begin{equation}
    \mathrm{Error} \propto \frac{a}{N^b}
\end{equation}

\noindent Learning curves of functional machine learning models thus {\em must} behave linearly on log-log scales, and form, in the limit of large $N$, a robust tool to assess learning data-efficiency through comparison of the  ``offset,'' $\log\left(a\right)$, and ``slope,'' $b$.

This paper is structured as follows: we first describe our methodology and illustrate the idea of learning curve comparison  for a well-known toy system from mathematics, the well-described Himmelblau's function, in order to demonstrate how different types of training data affect the learning rate for a generic non-trivial 2-dimensional surface.\cite{himmelblau1972}
Next, we employ the same testing scheme for two different but common use cases in chemistry:
The first use case concerns generating a force field for a single molecule. 
Here we demonstrate how using forces and energies in the training data affect the learning rate for the potential energy surfaces of 10 individual molecules from the MD17 dataset.\cite{chmiela2017machine,Chmiela2019sGDML}
In order to ensure that the underlying data is practically noise-free, we additionally present a revision of the MD17 where the energies and forces have been calculated using dense integration grids and tight self-consistent field (SCF) tolerances.

The second use case concerns generating machine learning models trained {\em across} chemical compound space, meaning that they can extrapolate to out-of-sample molecules not seen during training. 
Also here, we investigate how the inclusion of forces in the training data helps the trained machine to generalize from points on the potential energy surfaces of known molecules to points on the potential energy surfaces of new molecules.\\
%

\section{Theory}
In order to achieve a fair comparison between models trained on different kinds of data (here energies and forces), we employ three closely related kernel-based regression models which are described in this section, as well as standard training protocols~\cite{AssessmentMLJCTC2013}.
Using these schemes, we ensure that models trained on different types of data will rely on the same kernel functions and representations while at the same time being exactly-determined with closed-form solutions.
This is done by defining loss functions in which basis kernel functions are placed on the training data, as well as placing kernel derivatives corresponding to any gradient or force information in the training data.
This yields a set of basis functions which guarantees that the model (within a given regularization) is able to perfectly align with the training labels. 
%
Additionally, use of the force operator ensures that the resulting force fields are conservative force fields, which is important for molecular dynamics simulations, and that the predicted energies obey common physical relations such as rotational and translational invariances.

The three machine learning models which employ training on either energies (function) only, or  energies and forces (function and derivatives) simultaneously, or forces (derivatives) only are presented next. 
Each model is characterized by its loss function $J$ which is minimized through regression. Note, that prediction errors can also be evaluated using different loss functions. 
Conventionally, training and test error definition are identical,  and their evaluation only differs in being assessed on a training or test set, respectively.
As such, there are nine possibilities to combine the three loss functions in training and in testing. 

While the relevant equations for each regressor can be derived both in the context of kernel ridge regression (KRR), as well as Gaussian process regression (GPR), we present them here in the notation most commonly used for the former.\cite{RasmussenWilliams,Glielmo2018,bpkc2010,Zhenwei2015,chmiela2017machine,Chmiela2019sGDML,christensen2018operators,Christensen2020FCHL19}
For more detailed derivations, we refer to the work by Bart\'ok and Cs\'anyi,\cite{GAPtutorial} as well as that of Mathias.\cite{sonjamathias}
We further note that the equations for the force-only regressor have recently seen a different derivation in the work on ``gradient domain'' machine learning (GDML) by Chmiela et al.\cite{chmiela2017machine}

\subsection{Energy-only Training}
In kernel ridge regression, the energy, $u$, of a query molecule, represented by $\tilde{M}$, can be expanded using a basis set of kernel functions placed on training samples.
That is,
\begin{equation}
        u\left(\tilde{M}\right) = \sum_i^\text{training} \mathpzc{k}\left(\tilde{M}, M_i \right)\alpha_i,
\end{equation}
where $M_i$ is the $i$-th molecule in the training set, $\alpha_i$ is the $i$-th regression coefficient with units of energy, and $\mathpzc{k}\left(\cdot,\cdot\right)$ is a function that relates two molecules through a similarity measure which typically depends on the specific functional form of the kernel function, the representation, and the metric.
Writing the above equation on matrix form yields:
\begin{equation}
        \mathbf{u} = \mathbf{K} \bm{\alpha}\label{eq:krr}
\end{equation}
Here, $\mathbf{u}$ is the vector containing predicted energies, $\mathbf{K}$ is the kernel matrix containing the pairwise kernel elements between the molecules in the training set and the predictions sets, and finally $\bm{\alpha}$ is the vector containing regression coefficients.
Bold lowercase letters indicate vectors and bold capital letters indicate matrices.
The specific kernel function used for the different experiments are given in Section \ref{sec:kernelfunctions}.
From Eq.~\ref{eq:krr}, a set of regression coefficients can be obtained by minimizing a loss function.
Using Thikonov regularization and minimizing the squared errors leads to the loss function which is the foundation of KRR:
\begin{equation} \label{eq:krr_lagrangian}
    J(\bm{\alpha}  \mid \{ \mathbf{u}^\mathrm{ref}\} ) = \tfrac{1}{2}\norm{ \mathbf{K} \bm{\alpha} - \mathbf{u}^\mathrm{ref} }_2^2 
        + \frac{\lambda}{2} \bm{\alpha}^\top \mathbf{K} \bm{\alpha} 
\end{equation}
This loss function contains a hyperparameter, $\lambda$, which determines the amount of L2-regularization for the regression coefficients stored in $\bm{\alpha}$, which in turn yields the well-known closed-form solution for training a KRR model:
\begin{equation}
         \bm{\alpha}  =  \left( \mathbf{K} + \mathbf{I} \lambda \right)^{-1} \mathbf{u}^\mathrm{ref}.\label{eq_krr_solution}
\end{equation}
These regression coefficients can then be used to predict energies (as in Eq.~\ref{eq:krr}), through evaluation of the relevant kernel matrix.
%
%
Via the definition of the force operator, i.e. the negative gradient of the energy with respect to the atomic coordinate vector, $\mathbf{r}$, the atomic force vector, $\mathbf{f}$, can be predicted by differentiation of Eq.~\ref{eq:krr}:
\begin{equation}
    \mathbf{f} \triangleq -\tfrac{\partial}{\partial\mathbf{r}} \mathbf{u} = \begin{bmatrix} -\frac{\partial}{\partial \mathbf{r}}\mathbf{K} \end{bmatrix}   \bm{\alpha}
\end{equation}
Note that this approach is as `naive' as general, and in principle any differential property can be learned this way as long as the representation accounts for all the variables which are being perturbed.\cite{christensen2018operators}
%
    
\subsection{Combined Energy and Force Training}\label{sec:gpr}
The second model investigated herein, is a model closely related to conventional KRR but which allows for training on both force and energies simultaneously.
Similar to how KRR can be used to construct a function that goes through the training points exactly, it is also possible to---at the same time---enforce the derivatives at those points.
Here, derivatives are enforced by including the derivatives (for example, forces) in the regression in addition to the function value (for example, energies).
In order to allow the model to match both the function values and derivatives of the training set exactly, it is necessary to increase the number of basis functions considerably, as the set of equations would otherwise be vastly overdetermined, as, for example, there are $3N$ force labels for each energy label.
One choice of extended basis comes from augmenting the set of basis functions in standard KRR with additional kernel functions corresponding to the kernel derivatives with respect to the coordinates of the training molecules.

The resulting set of equations for this problem looks as follows:
\begin{equation}\label{eq:gpr_derivative}
    \begin{bmatrix}
        \mathbf{u} \\
        \mathbf{f} 
    \end{bmatrix} = \begin{bmatrix}
       \mathbf{K} && -\frac{\partial}{\partial \mathbf{r}^\intercal}\mathbf{K} \\
       -\frac{\partial}{\partial \mathbf{r}}\mathbf{K} && \frac{\partial^2}{\partial \mathbf{r}\partial \mathbf{r}^\intercal}\mathbf{K} 
    \end{bmatrix} \bm{\alpha}
\end{equation}
Here, $\mathbf{K}$ is again the kernel matrix containing the pairwise kernel elements between the molecules in the training set and the predictions sets, and is defined identially to $\mathbf{K}$ in Eq.~\ref{eq:krr}.
Similarly to the problem in Eq.~\ref{eq:krr}, this choice of basis functions ensures that the training kernel is square-symmetric and the  minimization of the regularized loss function has a convenient closed-form solution.
For a set of reference training energies and forces, the regression coefficients can here be obtained by minimizing the following set of squared errors with Thikonov regularization:
\begin{equation} \label{eq:gpr_lagrangian}
J(\bm{\alpha} \mid \{ \mathbf{u}^\mathrm{ref}, \mathbf{f}^\mathrm{ref}\}) = \frac{1}{2}\norm{ \begin{bmatrix}
       \mathbf{K} && -\frac{\partial}{\partial \mathbf{r}^\intercal}\mathbf{K} \\
       -\frac{\partial}{\partial \mathbf{r}}\mathbf{K} && \frac{\partial^2}{\partial \mathbf{r}\partial \mathbf{r}^\intercal}\mathbf{K} 
    \end{bmatrix}  \bm{\alpha} -  \begin{bmatrix}
        \mathbf{u}^\mathrm{ref} \\
        \mathbf{f}^\mathrm{ref} 
    \end{bmatrix}  }_2^2 
    + \frac{\lambda}{2} \bm{\alpha}^\top  \begin{bmatrix}
       \mathbf{K} && -\frac{\partial}{\partial \mathbf{r}^\intercal}\mathbf{K} \\
       -\frac{\partial}{\partial \mathbf{r}}\mathbf{K} && \frac{\partial^2}{\partial \mathbf{r}\partial \mathbf{r}^\intercal}\mathbf{K} 
    \end{bmatrix}  \bm{\alpha} 
\end{equation}
The closed-form solution is, similarly to that of Eq.~\ref{eq_krr_solution},
\begin{equation} \label{eq:gpr_solve}
         \bm{\alpha}  =  \Biggl(  \begin{bmatrix}
           \mathbf{K} && -\frac{\partial}{\partial \mathbf{r}^\intercal}\mathbf{K} \\
           -\frac{\partial}{\partial \mathbf{r}}\mathbf{K} && \frac{\partial^2}{\partial \mathbf{r}\partial \mathbf{r}^\intercal}\mathbf{K} 
        \end{bmatrix} + \mathbf{I} \lambda \Biggr)^{-1} \begin{bmatrix}
            \mathbf{u}^\mathrm{ref} \\
            \mathbf{f}^\mathrm{ref} 
        \end{bmatrix}.
\end{equation}
This definition guarantees that the regression problem is exactly determined, and allows for the machine-learned potential energy surface to match both the energy and the forces of each training molecule.

With the regression coefficients obtained through Eq.~\ref{eq:gpr_solve}, predicted energies can then be calculated via the set of kernel functions and kernel derivatives placed on the training set:
\begin{equation}\label{eq:gpr_energy}
    \mathbf{u} = 
    \begin{bmatrix}
        \mathbf{K} \quad -\frac{\partial}{\partial \mathbf{r}^\intercal}\mathbf{K}
    \end{bmatrix} \bm{\alpha}
\end{equation}
Likewise, forces can be obtained by taking the relevant derivative of the energy, i.e.:
\begin{equation}\label{eq:gpr_force}
        \mathbf{f} \triangleq -\tfrac{\partial}{\partial\mathbf{r}} \mathbf{u} = \begin{bmatrix}
       -\frac{\partial}{\partial \mathbf{r}}\mathbf{K} && \frac{\partial^2}{\partial \mathbf{r}\partial \mathbf{r}^\intercal}\mathbf{K} 
    \end{bmatrix} \bm{\alpha}
\end{equation}

\subsection{Force-only Training}
The third model presented is a model which constructs a potential energy surface from information about derivatives only.
This model is similar to the model described in Section~\ref{sec:gpr} except that, in this case, only force labels are used in the training step, and, consequently, the set of basis functions is comprised of only the corresponding kernel derivatives.

In this approach, we start with the following equation, in which the kernel matrix is formed by the double derivatives of pair-wise kernel functions:
\begin{equation}
    \mathbf{f} =         \begin{bmatrix}  
    \frac{\partial^2}{\partial \mathbf{r}\partial \mathbf{r}^\intercal}\mathbf{K}
        \end{bmatrix}  \label{eq:gdml_force}
    \bm{\alpha}
\end{equation}
For a set of reference force labels, $\mathbf{f}^\mathrm{ref}$, the corresponding regression coefficients are obtained analogously to the energy-only and energy+force examples in the previous sections. 
\begin{equation} \label{eq:gdml_lagrangian}
J(\bm{\alpha}\mid \{\mathbf{f}^\mathrm{ref}\}) = \tfrac{1}{2}\norm{ \begin{bmatrix}  
    \frac{\partial^2}{\partial \mathbf{r}\partial \mathbf{r}^\intercal}\mathbf{K}
        \end{bmatrix}  \bm{\alpha} - \mathbf{f}^\mathrm{ref} }_2^2 
    + \tfrac{\lambda}{2} \bm{\alpha}^\top \begin{bmatrix}  
    \frac{\partial^2}{\partial \mathbf{r}\partial \mathbf{r}^\intercal}\mathbf{K}
        \end{bmatrix}  \bm{\alpha} 
\end{equation}

\begin{equation}
         \bm{\alpha}  =  \biggl(  \begin{bmatrix}  
        \frac{\partial^2}{\partial \mathbf{r}\partial \mathbf{r}^\intercal}\mathbf{K}
            \end{bmatrix} + \mathbf{I} \lambda \biggl)^{-1} \mathbf{f}^\mathrm{ref}.
\end{equation}
This is also sometimes referred to as training in the ``gradient domain''.\cite{chmiela2017machine}

Energies can then be predicted through direct integration of Eq.~\ref{eq:gdml_force}, which allows the scalar field to be determined up to an integration constant:
    \begin{equation}
        \mathbf{u} = \begin{bmatrix} -\frac{\partial}{\partial \mathbf{r}^\intercal}\mathbf{K} \end{bmatrix}   \bm{\alpha} + \mathrm{constant~term}
    \end{equation}
For problems where only a single surface is of interest---for example the potential energy surface of a given stoichiometry---the integration constant is rarely of any importance, and can also be inferred by predicting the energies for the training data.
Unfortunately, however, this approach is less practical when multiple surfaces are of interest, for example for datasets involving the potential energy surfaces of multiple different molecules.
Note that the number of possible stoichiometries is known to grow exponentially with elementary particle number~\cite{anatole-ijqc2013}.
Hence, force-only training makes it hard if not impossible to train models that directly predict energies for molecules of varying size and chemical composition, if the energy of the molecules are of interest.
Of course, composite approaches, e.g.~using dressed atom models~\cite{BOB}, can still be used to rectify such shortcomings. 

\subsection{Representations and Kernel Functions}\label{sec:kernelfunctions}
Since we present machine learning models trained on (i) an analytical function (Himmelblau's function) and  on  (ii) several models of molecular energetics trained non-equilibrium geometries, different kernel and representation choices have been made.
\subsubsection{Representing Himmelblau's Function}
Here, Himmelblau's function is used as a toy-model for learning complicated surfaces.\cite{himmelblau1972}
This allows for thorough benchmarking of the regressors used herein.
Himmelblau's function is a multi-modal function with one local maximum and four minima and bears some resemblance to the potential energy surface of a simple molecule with four conformational minimia.
The function is defined as:
\begin{equation}
u\left(x, y\right) = \left( x^2 + y - 11\right)^2 + \left(x + y^2 -7\right)^2.
\end{equation}
Points on the surface are represented here as their $xy$-coordinate pair, that is,~$\mathbf{q}_i = \left[ x_i, y_i \right]^{\top}$.
A Gaussian kernel function is used to compute the kernel elements for Himmelblau's function, i.e.:
\begin{equation}\label{eq:kernel_himmelblau}
     \mathbf{K}_{ij} = \exp\left(-\frac{\| \mathbf{q}_i - \mathbf{q}_j \|^2_2}{2\sigma^2}\right) \quad \text{for Himmelblau's function}
\end{equation}
%
This choice of representation and kernel function makes is straightforward to implement and enables the computation of necessary kernels and derivatives analytically.

\subsubsection{Representing Molecules}
To represent the environments of an atom in a molecule, we rely on the computationally more efficient variant of the Faber-Christensen-Huang-Lilienfeld (FCHL) representation\cite{FCHL}, namely the FCHL19 representation \cite{Christensen2020FCHL19}.
Briefly, this representation is a vector which contains histograms of the radial distributions of atoms and a number of Fourier terms describing angular distributions of atoms in the environment of a certain atom.\cite{faber2017alchemical}
In principle, any continuous representation that generalizes across chemical space could have been used for this purpose.
%
In addition, we use the localized kernel \textit{ansatz} in which the kernel elements between two molecules correspond to the pairwise sum over the kernel functions between the respective representations of atomic environments in the two molecules.\cite{bpkc2010,barker2016lcgap}
This makes it possible to train models that span molecules of varying size and chemical composition.
The following Gaussian kernel function is used throughout:
\begin{equation}\label{eq:kernel_screening}
     \mathbf{K}_{ij} =  \sum_{I \in i}  \sum_{J \in i} \delta_{Z_I Z_{J}}  \exp\left(-\frac{\| \mathbf{q}_I - \mathbf{q}_{J} \|^2_2}{2\sigma^2}\right) \quad \text{for molecules} 
\end{equation}
%
where $\mathbf{q}_I$ and $\mathbf{q}_{J}$ are the representations of the $I$'th and $J$'th atoms in the molecules $i$ and $j$, respectively, $\delta$ is the Kronecker delta, with $Z_I$ and $Z_J$ being the atomic numbers of each atom, respectively.

\section{Results}

In this section, we present numerical evidence which demonstrates the effects on the learning rate of including derivative labels in the training data.
This section is organized as follows: first, we establish the generality of our numerical experiment by learning the surface of a simple two-dimensional function, unrelated to molecules or chemistry.
Next, we demonstrate the same principles applied to two distinct use cases, namely (1) training a model for the potential energy surface for a single molecule, and (2) training a general model for the potential energy surfaces of a number of molecules of different size and chemical composition.

\subsection{Toy system: Learning Himmelblau's Function}
\begin{figure}
    \centering
    \includegraphics[width=\linewidth]{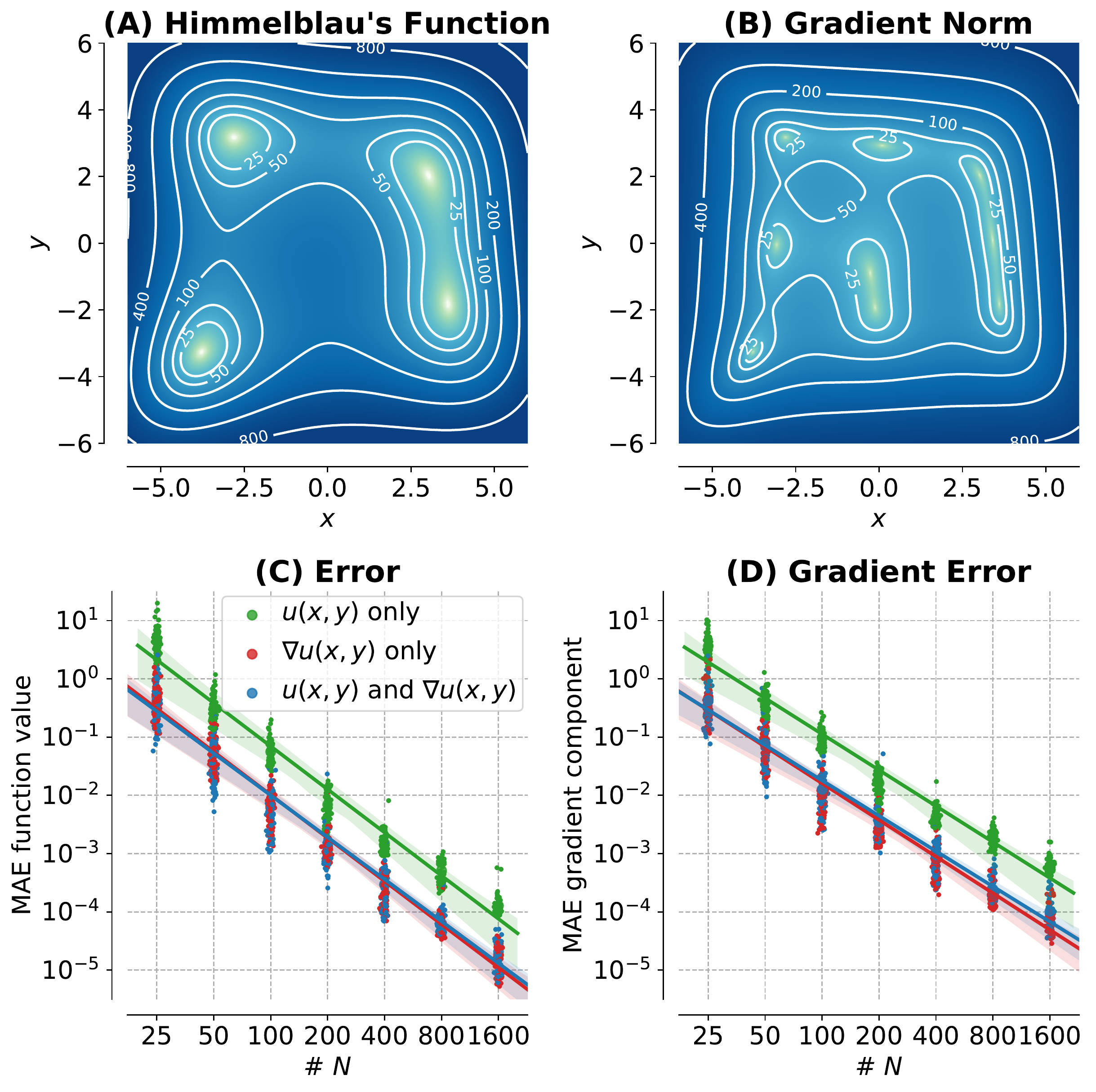}
    \caption{\label{fig:himmelblau} Panel (A) Displays the surface of Himmelblau's function, and (B) the gradient norm of the surface. Panels (C) and (D) display the MAE prediction and MAE gradient prediciton, respectively, of the surface of Himmelblau's function as a function of the training set size. Three different sets of training labels are used: green is trained on only the function values, red is trained only on function gradient components, and finally blue is trained using both simultaneously.}
\end{figure}

In this section we investigate the learning curves of machines trained to predict Himmelblau's Function.
This 2D surface has four local minima and is displayed in Fig.~\ref{fig:himmelblau}, as well as the norm of its gradient.
Here, Himmelblau's function serves as a toy system to demonstrate the effects of including function derivatives in the training procedure.

We generate a dataset by selecting random points from the surface.
The test set consists of 10,000 points sampled randomly and uniformly across the interval $ \{x_i, y_i\} \in \left[-6.0; 6.0 \right]$, with training sets of varying sizes sampled randomly and uniformly across the same interval.
For each training set size $N \in \{25, 50, 100, 200, 400, 800, 1,600\}$, 100 training and test sets are created using 100-fold random sub-sampling cross-validation with a constant test set size of 10,000 points.
This extensive cross-validation was necessary in order to have well-converged averages over the folds, as the test errors were observed to vary up to one order of magnitude.
In total over 350,000 individual machines were trained in order to obtain the presented learning curves.

Three different models were trained for the surface: the first model trained on only the function values, the second only trained on the function derivatives at the training points, and the third model was trained on both the function values and function derivatives simultaneously.
As expected,\cite{vapnik1994learningcurves,StatError_Muller1996} the three machines yield learning curve fits for the predicted mean-absolute-error (MAE) function value with similar slopes when plotted on a log-log scale, as displayed in Fig.~\ref{fig:himmelblau}C.

The average difference between the two models which include forces labels in the training set is much smaller than the standard deviation of the 100 folds used to calculate the average.
Compared to training on function values exclusively, the average decrease in the off-set of the learning-curve was found to be a factor of 7.0 for the model trained on only function derivatives, and 7.2 for the model trained on both function values and function derivatives.

\begin{table}
     \centering
     \caption{Slopes and offsets for predictions of the MAE value and MAE gradient component for Himmelblau's function, based on three machine learning methods trained on different types of data. \label{tab:slopes_himmelblau}}
  \begin{tabular}{llrrrr}
    \bottomrule
                           & \multicolumn{2}{l}{MAE Value} & \multicolumn{2}{l}{MAE Gradient}  \\
        Training data          & Slope  & Offset & Slope  & Offset \\
        \midrule
        Values             & -2.45  & 8.49 & -2.05  & 7.14 \\  
        Values and gradients   & -2.33  & 5.81 & -1.96  & 4.76 \\  
        Gradients          & -2.43  & 6.43 & -2.06  & 5.16 \\  
        \bottomrule
      \end{tabular}
    \end{table}

Learning curves for gradient predictions show similar trends among the three models (see Fig.~\ref{fig:himmelblau}D):
the average decrease in the offset was found to be a factor of 7.9, when training on gradients only, and 7.0 when training on both function values and gradients simultaneously, compared to training only on function values.

Fitted slopes and offsets for all six learning curves are displayed in Table \ref{tab:slopes_himmelblau}.
We note that the 95\% confidence intervals of the fitted learning curves for the two models that include gradients in the training loss-function are mostly overlapping, and while slopes and offsets differ somewhat (see Table \ref{tab:slopes_himmelblau}), the difference between the two models is not statistically significant.

\begin{table}
\footnotesize
     \centering
     \caption{Learning curves for three machines trained on different types of data, and the same loss functions evaluated on a test set of 10,000 points on the surface of Himmelblau's function. The three machines are trained on either function values, function values and gradients simultaneously, or gradients only. The loss functions used to train each machine is denoted in the left column, followed by the slopes and offsets for the resulting learning curves for three loss functions evaulated on the test data. The three loss functions are defined in the text in Eqn.~\ref{eq:krr_lagrangian},~\ref{eq:gpr_lagrangian}, and~\ref{eq:gdml_lagrangian}, respectively. \label{tab:cost_functions} }
  \begin{tabular}{llrrrrrr}
  \toprule
Training & & \multicolumn{2}{l}{Test loss} & \multicolumn{2}{l}{Test loss} & \multicolumn{2}{l}{Test loss} \\   
 \cmidrule(lr){1-2}\cmidrule(lr){3-4}\cmidrule(lr){5-6}\cmidrule(lr){7-8}

&  & \multicolumn{2}{l}{$J(\bm{\alpha} \mid \{ \mathbf{u}^\mathrm{test}\})$}   
  & \multicolumn{2}{l}{$J(\bm{\alpha} \mid \{ \mathbf{u}^\mathrm{test}, \mathbf{f}^\mathrm{test}\})$}   
  & \multicolumn{2}{l}{$J(\bm{\alpha} \mid \{ \mathbf{f}^\mathrm{test}\})$} \\
Training data &  Training loss     &   Slope & Offset     &   Slope & Offset     &   Slope & Offset \\
   \midrule/
Values  &$J(\bm{\alpha} \mid \{ \mathbf{u}^\mathrm{ref}\})$ &	-4.59 &	27.06	&   -3.74 &	25.19 & -3.63 &	24.41 \\
Values and gradients &$J(\bm{\alpha} \mid \{ \mathbf{u}^\mathrm{ref}, \mathbf{f}^\mathrm{ref}\})$ &	-4.41 &	22.35	&   -3.58 &	20.80 & -3.49 &	20.25 \\
Gradients  &$J(\bm{\alpha} \mid \{ \mathbf{f}^\mathrm{ref}\})$ &	-4.60 &	23.10	&   -3.78 &	21.51 & -3.69 &	20.91 \\
        \bottomrule
      \end{tabular}
    \end{table} 

Also for learning curves using loss functions we find agreement with the power-law behavior that is expected from models trained on function values,\cite{vapnik1994learningcurves,StatError_Muller1996} here demonstrated for models trained on function gradients.
In Table~\ref{tab:cost_functions}, we present resulting slopes and offsets for loss-function learning curves for the three types of trained machines trained using three different sets of training data and corresponding loss functions (rows).
These learning curves are shown graphically in Fig.~\ref{fig:lc_fit_cost} in the Supplementary Information.

Results for the corresponding three test loss functions (columns) are obtained using a test set of 10,000 points randomly selected from the surface of Himmelblau's function.
Here, the same trends as for the MAE learning curves are observed. 
The resulting slopes from the three types of trained machines do not differ with statistical significance, but instead depend on the type of test data.
In contrast, the offsets depend strongly on the training loss function and training data.
More specifically, the inclusion of gradients in the training loss function is on average beneficial when predicting function values, function and gradient values, as well as gradient values alone.
We also remind the reader that the slope of these learning curves are independent of the units of the labels, as this quantity is folded into the offset.
Despite using hundred-fold cross-validation, we do not observe a statistically significant difference between any of the three test predictions when using training loss functions and data sets based on function values and gradients, or gradients alone. 
We also note that the learning curves for the MAE error are observed to follow the expected power-law behavior over a range that spans 6 and 5 orders of magnitudes for predicted function values and predicted function gradient components, respectively.

We have thus demonstrated for the Himmelblau function, that MAE error in this case can be decreased 7-fold by inclusion of labels that correspond to function derivatives in the training set.
At the same time, however, we find negligible improvement upon inclusion of function values, in addition to the gradient information in the training algorithm.
Interestingly, these results suggest that when learning a single function surface, using derivatives as training labels is more advantageous than actual function values.

\subsection{Use case 1: Learning the PES for a one molecule}

    \begin{figure}
        \centering
        \includegraphics[width=\linewidth]{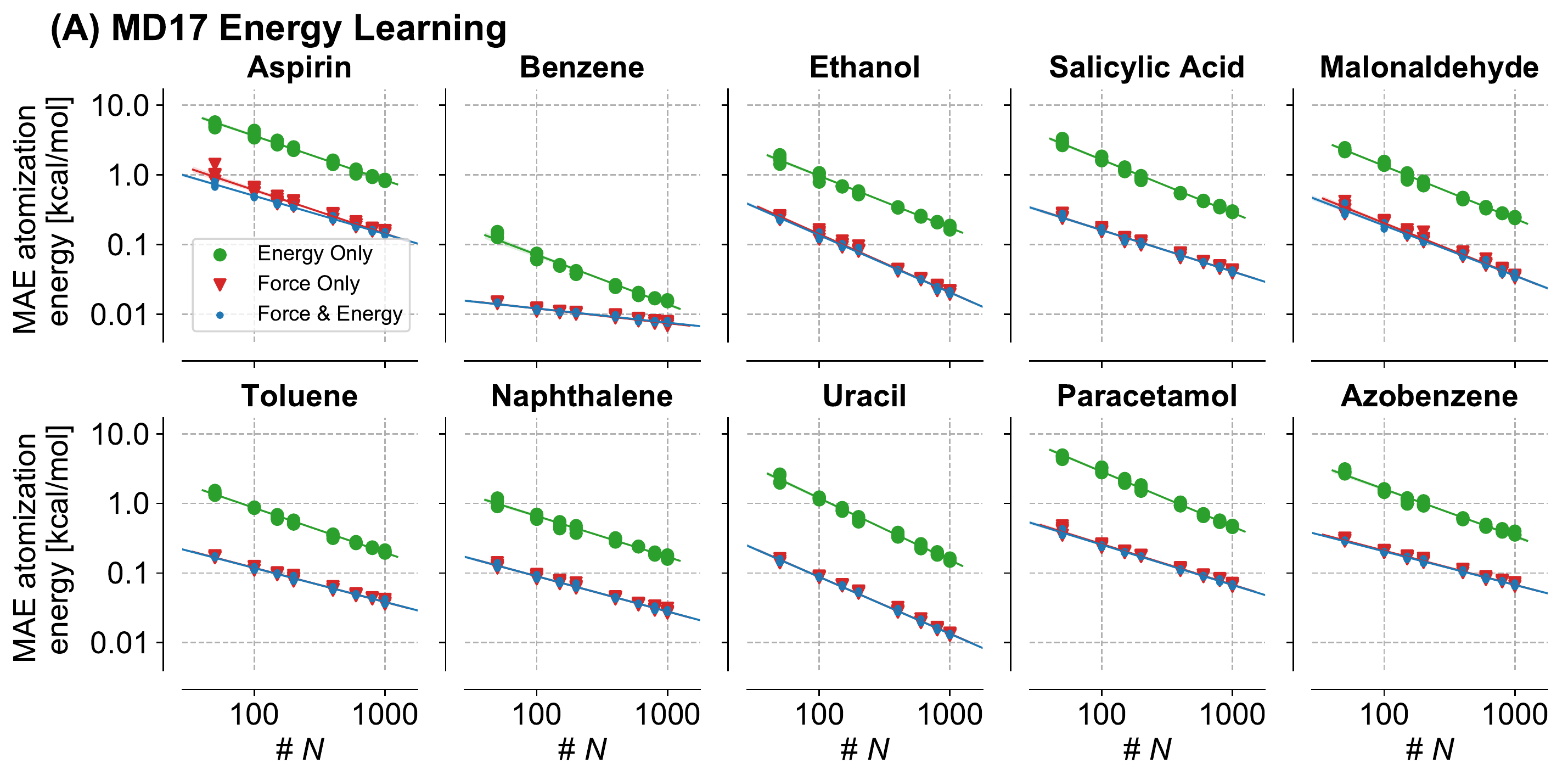}
        \includegraphics[width=\linewidth]{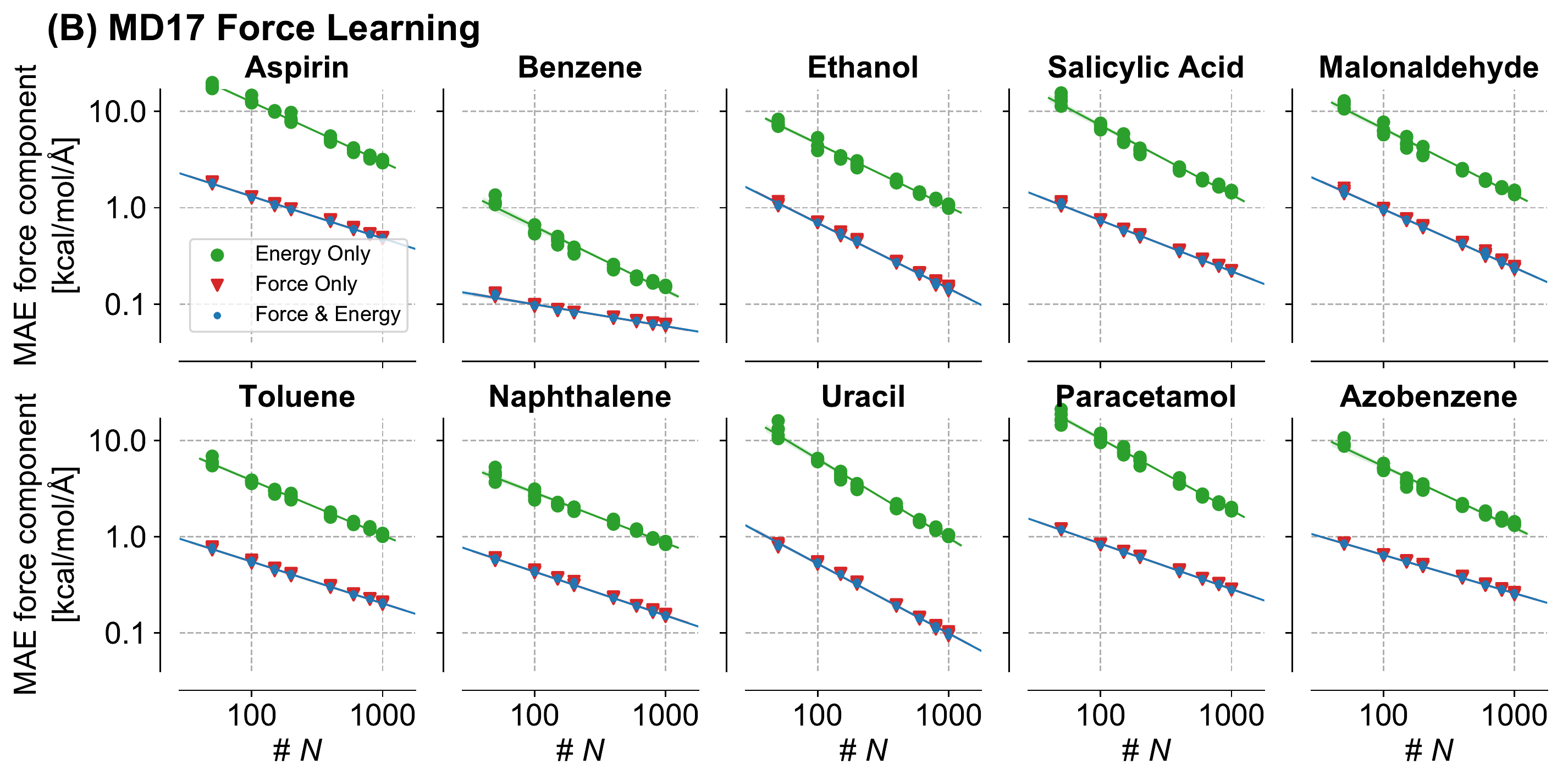}
        \caption{Learning curves for machines trained on three different datatypes for the 10 molecules in the revised MD17 dataset, namely machines trained on energies only, forces only, and both forces and energies. 
        Data for five folds are plotted as a scatter plot for each training set size, and a linear fit for each curve is plotted in addition.
        In (A) the mean absolute error (MAE) of the predicted atomization energy is presented for each molecule as a function of the number of molecules in the training set, while (B) shows MAE of the predicted force components for the same folds.\label{fig:md17_fits}
            }
    \end{figure}
    
In this section we discuss the accuracy of energy and force predictions from three models trained on the revised MD17 dataset.\cite{chmiela2017machine,Chmiela2019sGDML}

For each of the 10 molecules in the revised MD17 dataset, three models are trained using either only the energy labels of each sample, only the force labels, or both types of labels simultaneously.
For each of the three models, the training set size $N$ is varied for $N \in \{ 50, 100, 150, 200, 400, 600, 800, 1000\}$.
%
%
The resulting MAE of energy and force predictions for each molecule can be found in Table \ref{tab:md17si} in the Supplementary Information, and plots with linear fits to the learning curves are shown in Fig.~\ref{fig:md17_fits}.
As it is clear from the results, and similarly to the example with Himmelblau's function, including the forces in the training procedure improves the prediction of both energies and forces.
On average, the MAE of the predicted energies and force components is reduced about 7-fold for the same number of training samples.
Regarding the inclusion of energies in addition to forces, we find, again in complete analogy to the Himmelblau function, that it makes little difference.
In all cases, the largest difference was found to be 0.001 kcal/mol MAE predicted energy, and 0.002 kcal/mol/\AA~MAE predicted force components at the largest training set size of $N$=1000 samples.

\begin{figure}
    \centering
    \includegraphics[width=\linewidth]{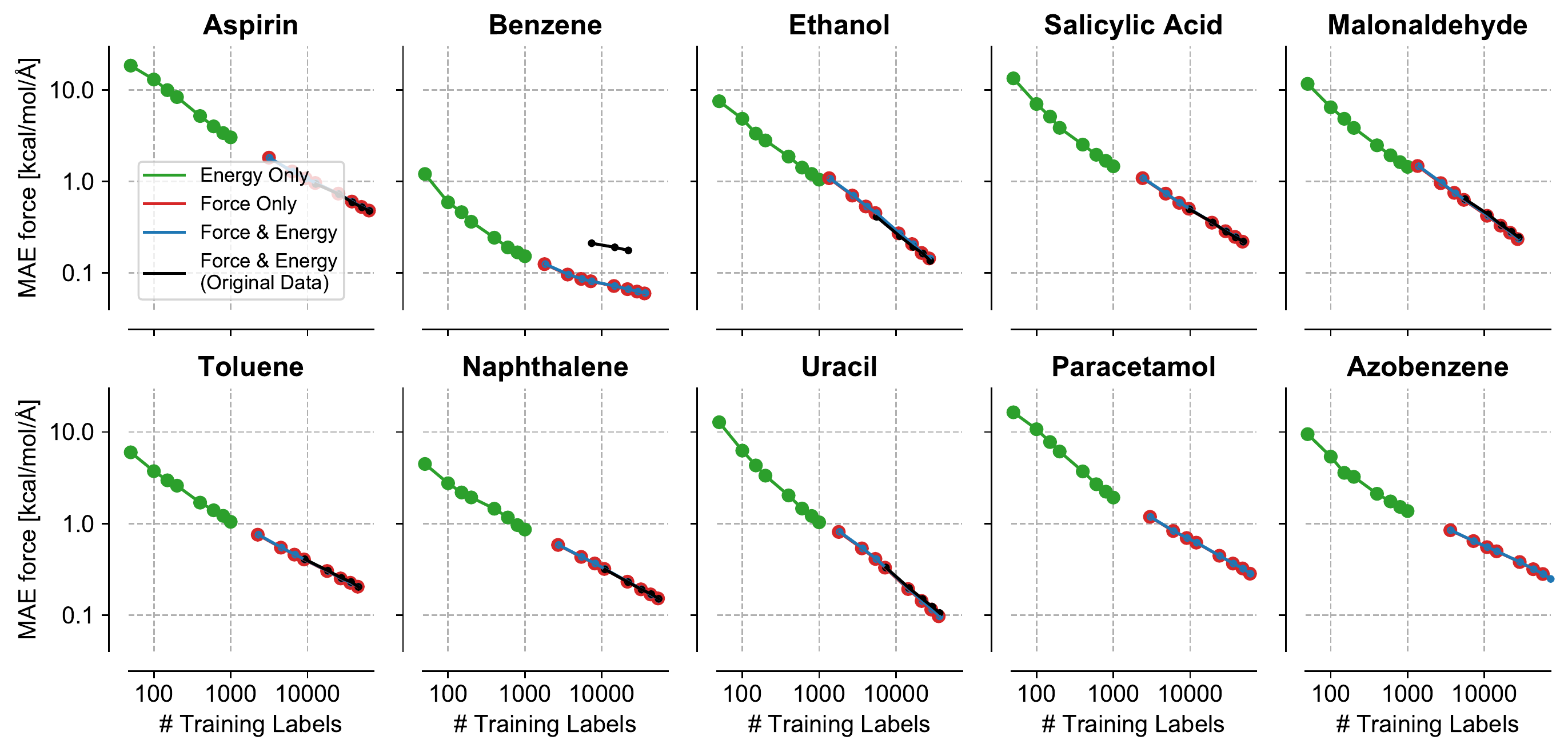}
    \includegraphics[width=\linewidth]{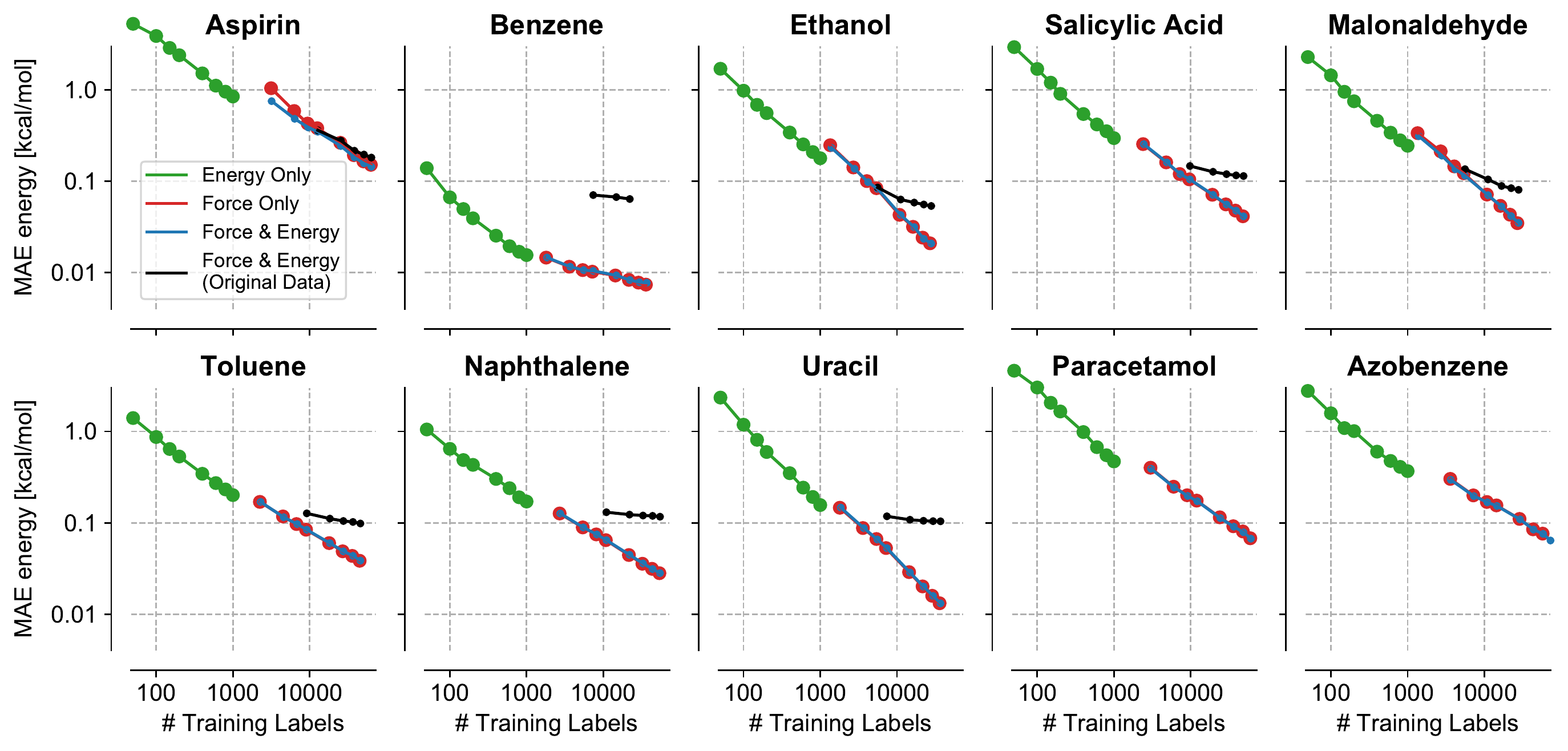}
    \caption{\label{fig:md17_labels} Force and energy learning curves for the revised MD17 dataset using various types of training data. In addition, learning curves for a machine trained on the original and somewhat noisier DFT data\cite{chmiela2017machine} is presented, marked as "Original Data".
        }
\end{figure}

In Fig.~\ref{fig:md17_labels} the MAE predicted energy is plotted as a function of the number of training labels for each of the 10 molecules using the three models described earlier.
Here, the slopes and offsets of the learning curve are close to identical in all 10 cases, regardless of what data was used to train the models.
This suggests that for this dataset, one force label yields as much improvement in the predictive accuracy of a machine learning model as one energy label.
Thus, a model for a specific molecule, trained on a certain number of energies, will have roughly the same predictive accuracy as one that is trained on the same number of force labels. 
As a consequence, in such cases---and if force-evaluations represent computationally negligible overhead (common within, for example, DFT)---the number of independent quantum calculations necessary to generate the training data required to reach a certain predictive power, can be reduced by a factor of 3$N$ thanks to the inclusion of forces.

We find that in all but one case, the learning curves for both force as well as energy prediction follows the expected power law, meaning they display a linear relationship on a log-log scale.
In the case of benzene, the off-set of the learning curve is extremely low, but the slope is inconsistent with the expected power-law.
It seems unlikely that underlying numerical noise in the training data is at the origin of such premature saturation: the learning curves for Uracil are equally low and do not suffer from lack of linearity.
The more likely culprit appears to be the specific parameterization of the FCHL19 representation achieving only insufficient uniqueness, a possibility recently pointed out by Pozdnyakov et al.\cite{pozdnyakov2020completeness}.
The addition of higher order terms, such as for example 4-body terms, into FCHL~\cite{FCHL} might rectify this issue. 
%
%
Furthermore, we note that the parameters in the representation have not been re-optimized for discriminating the subtle difference between the very similar structures in this, but rather work across molecules of varying sizes and chemical composition.
It is thus possible that they could be re-optimized to alleviate such issues to some extent.

\subsection{Use case 2: Training models across chemical composition}

\begin{figure}
    \centering
    \includegraphics[width=\linewidth]{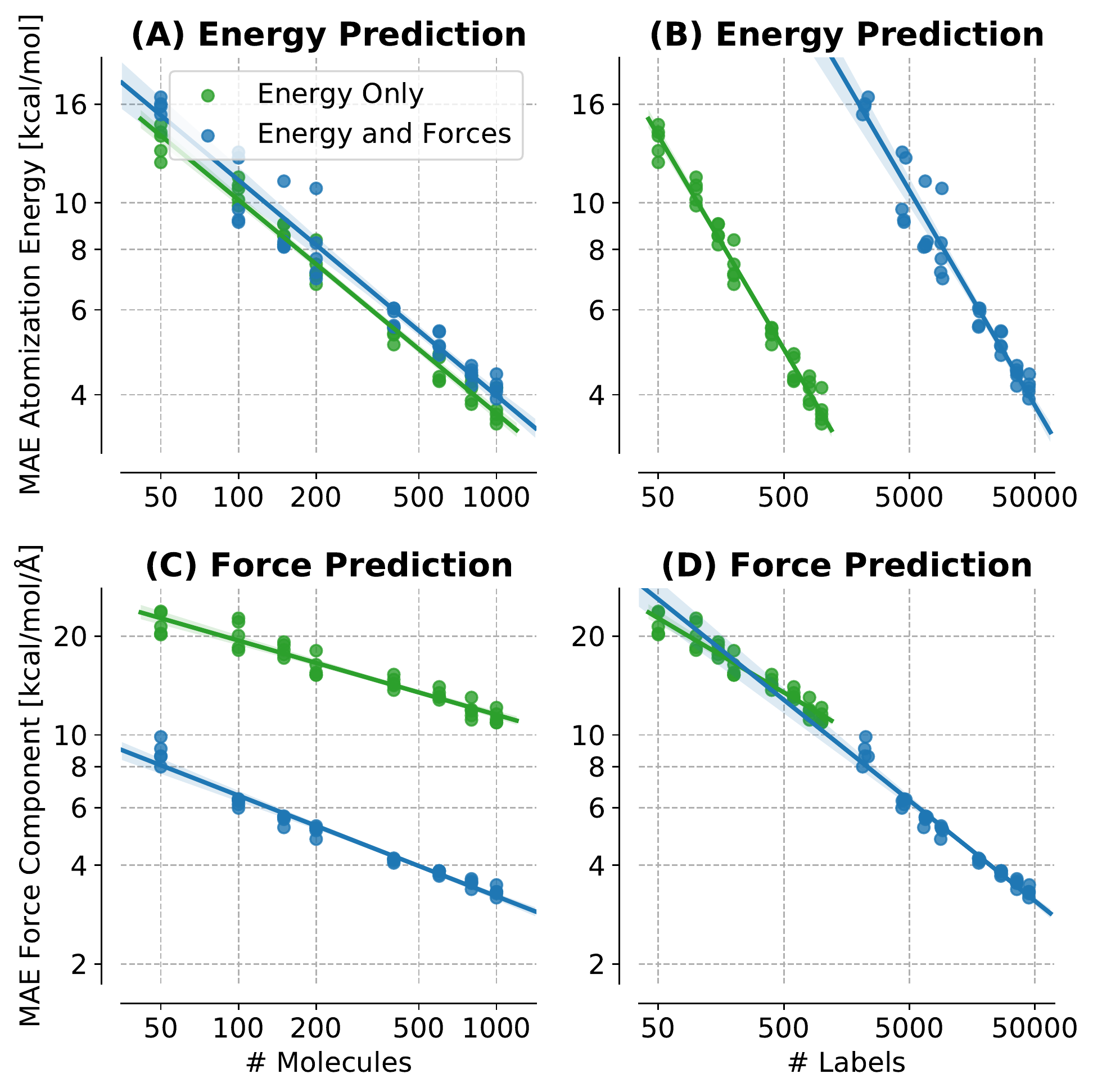}
    \caption{\label{fig:amons_learning} Energy and force learning curves machine learning models trained on a dataset consisting of non-equilibrium conformers for 1,595 small, organic molecules. 
    Two models are trained, one including only the energy labels of each non-equilibrium conformer, and one additionally including the corresponding force labels.
    Panel (A) shows the MAE predicted atomization energy as a function of the number of molecules in the training set, while (B) shows the same predicted quantity, but as a function of the total number of training labels (i.e.~the total number of energy and force components in the training set) for each machine.
    The MAE predicted force components are displayed in panels (C) and (D) as a function of the number of molecules in the training set (C) as well as the total number of training labels (D).}
\end{figure}

\begin{table}
    \footnotesize
     \centering
     \caption{Slope and offsets (see text) of linear fits to mean absolute error (MAE) of predicted energies and gradients versus the training set size, on a log-log scale. The dataset consists of 1,595 small organic molecules in nonequilibrium conformation (see text).  The column ``Training'' denotes what data was used to train the models. Lastly, the column ``$N$ unit'' denotes the unit is used for the abscissa of a given learning curve. [Molecules] denotes that the unit is simply the number of molecules in the training set, that is for energy-only training, one training label (the energy) is used per molecule, while for energy+forces, 3$N$+1 training labels (one energy and $3N$ force components) are used for each molecule, whereas [Labels] denotes that the total number of $3N+1$ labels are used on the abscissa.\label{tab:slopes_amons}}
  \begin{tabular}{llrrrr}
    \toprule
    &  & \multicolumn{2}{l}{MAE Energy [kcal/mol]} & \multicolumn{2}{l}{MAE Force [kcal/mol/\AA]}\\
    Training        &    $N$ unit &  Slope   & Offset & Slope & Offset\\
    \midrule
    Energy Only     & [Molecules] & -0.44  &  4.35 &  -0.23  &  4.01\\  
    Energy and Forces & [Molecules] & -0.45  &  4.47 &  -0.31  &  3.29\\  
    Energy and Forces & [Labels]    & -0.45  &  6.16 &  -0.31  &  4.46\\  
    \bottomrule
  \end{tabular}
\end{table} 

In our second use case scenario, we investigate how learning forces and energy across chemical compound space behaves as a function of the number of training molecules and the type of training data.
Fig.~\ref{fig:amons_learning} shows learning curves for force and energy predictions on a dataset consisting of non-equilibrium conformations of 1,595 small, organic molecules.
The training set is divided such that every (distorted) molecule of a given chemical composition is only seen once in either the training or test sets.
%
%
Using this data, two types of models are trained: one model is trained using only atomization energy labels, and a second model is trained using both atomization energy as well as the corresponding 3N components of the force vector.
As the model trained on only forces would only integrate the energy up to an arbitrary constant, the predicted energies would not be meaningful, and therefore this model is not used in this section.

Fig.~\ref{fig:amons_learning}A shows the MAE predicted atomization energy for the two models as a function of the number of molecules involved.
The learning rates hardly differ, with the largest deviation between the two curves being 3\% of the average MAE.
While this seems surprising in light of the effect of including forces for the MD17 dataset discussed in the previous section, this result has a simple explanation.
Since the gradient information only provides information about the underlying function up to an integration constant, the additional information is useless in providing information about the differences between the individual potential energy surface on which all the molecules are located.
In this case, this information is provided solely by the atomization energies and, in the end, the dominant error in the learned atomization energies comes from learning energy differences between the different constitutions of atoms in each molecule, and not from more subtle changes in molecular conformation.
The learning rates are displayed in Table \ref{tab:slopes_amons}, where the slopes and offsets for energy predictions are very close for the two different models.
For energy predictions, the slopes of the learning curves are -0.44 and -0.45 for the models trained on only energies and the models  simultaneously trained on energies and forces, respectively.
For force prediction, however, the slopes differ somewhat, at -0.23 and -0.31, for the same two models, respectively, indicating that the model trained on both energies and forces might be better at very large training sets.
We note, however, that the current training set is not large enough to confirm this, as this would require a force training set with at least 10,000 to 50,000 training molecules (See Fig.~\ref{fig:amons_learning}D), which, in turn, would render this experiment computationally intractable with a kernel matrix of up to around 2,000,000 $\times$ 2,000,000 elements.

When the curves are viewed as a function  of the total number of training labels (in Fig.~\ref{fig:amons_learning}B), the energy-only model requires about 51 times fewer labels to reach the same accuracy as the model trained on both forces and energies.
This indicates,
that the additional force labels do not aid the model in predicting energies of unseen molecules.
This is likely due to the fact that forces can only determine the potential surface up to an integration constant, which can only be learned using energy labels.

Next, in Fig.~\ref{fig:amons_learning}C, the MAE predicted force components are plotted for the two models as a function of the number of molecules in the training set.
On average, including the forces in the training set reduces the predicted MAE force component by a factor of 3.5.
Essentially, while the additional gradient information is not able to improve learning beyond the integration constant, it does help improve the prediction of the relative energy landscape of each potential energy surface, in turn leading to improved gradient predictions, but at the cost of much larger number of training labels.
Fig.~\ref{fig:amons_learning}D displays the same MAE predicted force components as in Fig.~\ref{fig:amons_learning}C, but as a function of the total number of training labels.
In this case, we observe that the two learning curves are very close-lying, similarly to what was found for the revised MD17 dataset in the previous section.
This suggests that for the training sizes investigated herein, one training energy label is worth roughly the same as one training force component label, although the differences in slopes indicate that the model trained on energies and forces might reach superior accuracy at much larger training sizes where the role of the integration constant is diminished.

\section{Methodology}
\subsection{Hyperparameter Selection and Learning Curves}

All learning curves were generated using nested cross-validation as implemented in \texttt{scikit-learn}\cite{scikit-learn} via the following recipe:
%
First, the datasets were randomized.
Secondly, the datasets were divided into 100 folds using random subsampling for Himmielblau's function, while datasets consisting of molecules were randomly divided into 5 folds using the \texttt{KFold} class implemented in \texttt{scikit-learn}.
Next, a grid-search with 4-fold cross-validation within the training set extracted from the fold was used to select the optimal choice of the hyperparameters $\sigma$ (the kernel width) and $\lambda$ (regularization strength), in order to avoid overfitting.
In order to select hyperparameters that simultaneously work well for both force and energy prediction, the following score function was used to select these:\cite{schutt2018schnet}
    \begin{equation}
       \mathcal{L}  = 0.01 \sum_i \left( U_i - \hat{U}_i \right)^2 + \sum_i \frac{1}{n_i} \| \mathbf{F}_i - \mathbf{\hat{F}}_i \|^2\label{eq:cost_function}
    \end{equation}
Here, $U_i$ is the energy of the $i$'th molecule, and $\mathbf{F}_i$ and $n_i$ are the force-vector and number of atoms in the same atom, respectively.
In the cases where either force/function derivatives or energies/function values were not included in the training data, the first or the second term was left out, respectively.

For the grid searches for  Himmelblau's function, the kernel width was tested in the range $\sigma \in \{0.4, 0.8, 1.6, 3.2, 6.4, 12.8, 25.6\}$, and the regularization strength was tested in the range $\lambda \in \{ 10^{-13}, 10^{-12}, 10^{-11}, 10^{-10}, 10^{-9}, 10^{-8}\}$.
For the molecular datasets (see next section), the grid-searches used values of $\sigma \in \{0.25, 0.5, 1.0, 2.0, 4.0, 8.0, 16.0, 32.0 \}$ for the kernel width, and $\lambda \in \{10^{-12}, 10^{-11}, 10^{-10}, 10^{-9}, 10^{-8}, 10^{-7}, 10^{-6}\}$ for the regularization strength.

\subsection{Software}
All machine learning models were implemented in \texttt{Python} using the \texttt{QML} machine learning package.\cite{qmlcode2017}.
The packages \texttt{Matplotlib} and \texttt{Seaborn} were used to plot all figures, and \texttt{Scikit-Learn} was used to obtain the linear fits to learning curves.\cite{hunter2007matplotlib,seaborn,scikit-learn}

\subsection{Datasets}
This subsection briefly presents the used datasets and their availability.

\subsubsection{Revised MD17 Dataset}
For each of the 10 molecules in the MD17 dataset\cite{chmiela2017machine,Chmiela2019sGDML} (aspirin, benzene, ethanol, salicylic acid, malonaldehyde, toluene, naphthalene, uracil, paracetamol, and azobenzene), 100,000 structures were randomly selected from the available MD trajectory data.
For each of these structures, a single-point force and energy evaluation was carried out at the DFT level.
All calculations were performed in ORCA 4.0.1, using the PBE functional and the def2-SVP basis set with the resolution-of-identity (RI) approximation for the Coulomb integrals.\cite{orca4,PBE,Weigend2005}
In order to have a minimal unsystematic error, the keyword \texttt{VeryTightSCF} was used to reduce the error from SCF convergence, while the keywords \texttt{Grid7} and \texttt{NoFinalGrid} were used to utilize the largest standard grid implemented in ORCA.
This data has been uploaded to \url{https://dx.doi.org/10.6084/m9.figshare.12672038} along with the indices used for the outer 5-fold cross-validation.

\subsubsection{Small organic molecules}    
\begin{figure}
    \centering
    \includegraphics[width=\linewidth]{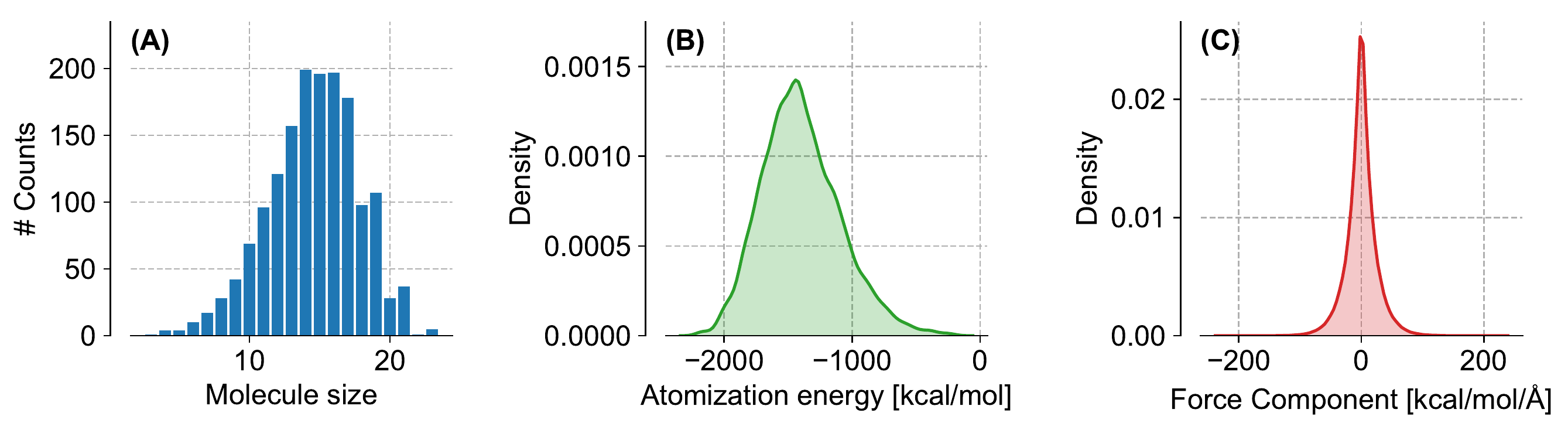}
    \caption{\label{fig:amons_data}The distribution of various properties of the dataset of small organic molecules.\cite{christensen2018operators} (A) shows the distribution of molecular sizes, while (B) shows the a kernel density estimate (KDE) plot of the distribution of atomization energies of the molecules, and lastly (C) shows the KDE plot of the force components in the dataset.}
\end{figure}
This dataset is taken from Ref.~\citen{christensen2018operators} and consists of non-equilibrium conformers of 1,595 small organic molecules with up to 7 atoms of the elements CNO saturated with hydrogen atoms.
This data is available for download at \url{https://dx.doi.org/10.6084/m9.figshare.7000280}.
For each of these structures generated via normal-mode sampling, the atomization energy and corresponding forces were provided at the $\omega$B97xD/6-31G(d) level of theory.\cite{Chai2008}
Fig.~\ref{fig:amons_data} shows plots of the distributions of molecular sizes, energies, and forces in the dataset.

\section{Conclusion and Outlook}
In this paper we have presented numerical confirmation that the predictive error of machines trained on function derivatives follows a power law, a result that is well-known and to be expected for both neural networks as well as kernel models trained on scalar values.
Our numerical results demonstrate this power-law behaviour for two popular and distinct types of molecular datasets, as well as for the Himmelblau's function for which
6 orders of magnitude have been spanned. 

For datasets that deal with a single surface, such as Himmelblaus's function and the potential energy surfaces of 10 molecules in the MD17 dataset, we find that including force labels in the set of training labels---in addition to energy labels---leads to an improvement of the predictive error when the same number of training points are used.
At the same time, however, we only find negligible differences between models trained on force labels only and models trained on both energy and force labels.
Regarding the number of total training labels (rather than number of molecules in the MD17 dataset), we found that the predictive error of forces and energies was close to identical, regardless of whether the model was trained on energies only, trained on both energies and forces, or trained on forces only.

For the diverse dataset of distinct organic molecules, we find that the prediction error of energies does not improve upon addition of force labels to energy labels in the training set, in contrast to what was observed for Himmelblau's function and the MD17 dataset.
Conversely, the predictive error of atomic forces is greatly improved by including forces in the training set.
The likely explanation is that geometrical derivative information only helps determine the surface up to an integration constant, corresponding to the atomization energy.
This effect is more visible when viewing the learning rate as a function of the number of training labels. In this case, energies are learned with about $20\times$ less training labels, training on energies only, compared to training on both forces and energies.
At the same time, the predictive force error when training on either energies, or forces and energies simultaneously, seems to decay at comparable rates with the number of force labels in these two scenarios.
In order to use derivative information to further improve this learning, it would be necessary to use \textit{alchemical} derivatives, i.e. the derivatives of the energy with respect to the atomic charges~\cite{anatole-prl2005,anatole-ijqc2013} which form the basis of alchemical perturbation density functional theory~\cite{guido2020apdft,guido2020alchemy}
These derivatives describe the change in energy as one molecule is alchemically transformed to another molecule, and could allow for better interpolation between potential energy surfaces of different molecules in a training set.

We believe that our observations have implications for the generation of new molecular datasets.
Here we point out that it is necessary to account for the cost of obtaining the training labels.
With this in mind, it is beneficial for use cases of the MD17-type to include forces labels in the training set when the cost of acquisition is less than the cost of $3N$ single-point energy calculations on un-correlated samples, where $N$ is the number of atoms in the molecule.
This suggests that for DFT-based datasets, it seems very favorable to calculate and report gradients in addition to single-point energies, while for methods with more costly gradients (compared to the energy evaluation) this becomes less favorable.

For energy predictions throughout large datasets used to fit models for general chemistry, i.e.~throughout chemical space, such as, for example, the family of ANI datasets\cite{smith2017ani,ANI_IsayevRoitberg2017,smith2018less,smith2020ani1cxx}, it seems more valuable to build the most compact model using a compositionally diverse training set with only single-point energies, rather than ``wasting'' coefficients by training also on forces for more conformations of the same molecule. 
If the goal, however, consists of {\em also} modeling forces, such as is typically the case for relaxing geometries throughout chemical compound space, our results indicate that the addition of forces (if acquisition cost is lower than for energies) in the training set is always beneficial. 
This conclusion, however, also depends on the requirements for execution speed and the availability of training data: Models trained on force labels can be computationally substantially more expensive to train and execute compared to models on the same number of energy labels, since this often involves the derivative of, for example, a kernel matrix or a neural network. 
Consequently, in an application scenario where sufficient energy labels are available, it might be best to train on energy labels only, as this enables numerically less complex training models.
Considering the prediction times for kernel-based models, it is also much more computationally expensive to evaluate kernel functions placed on derivatives compared to those placed on scalars.
If the ultimate goal is to have very fast prediction times for kernel-based models, it seems worthwhile to consider the use of kernel-based force models which do not require the evaluation of second-order kernel derivatives.

All these observations summarize our insights into the design of future data-driven models and their underlying dataset generation. We believe that they are applicable to all branches of machine learning where the goal is to learn multidimensional, differentiable function surfaces.

\section*{Acknowledgements}
This work was partly supported by the NCCR MARVEL, funded by the Swiss National Science Foundation.
O.A.v.L. acknowledges funding from the Swiss National Science foundation (407540\_167186 NFP 75 Big Data) and from the European Research Council (ERC-CoG grant QML). 
The authors thank Puck van Gerwen for helpful comments on the manuscript.
We further acknowledge the use of the following software: NumPy and the F2PY tool\cite{numpy}, and OpenMP.\cite{openmp08}

\section*{References}
\bibliographystyle{unsrt}
\bibliography{refs} 

\clearpage
\section*{Supplementary Material}
\clearpage
\beginsupplement
\begin{figure}
    \centering
    \includegraphics[width=\linewidth]{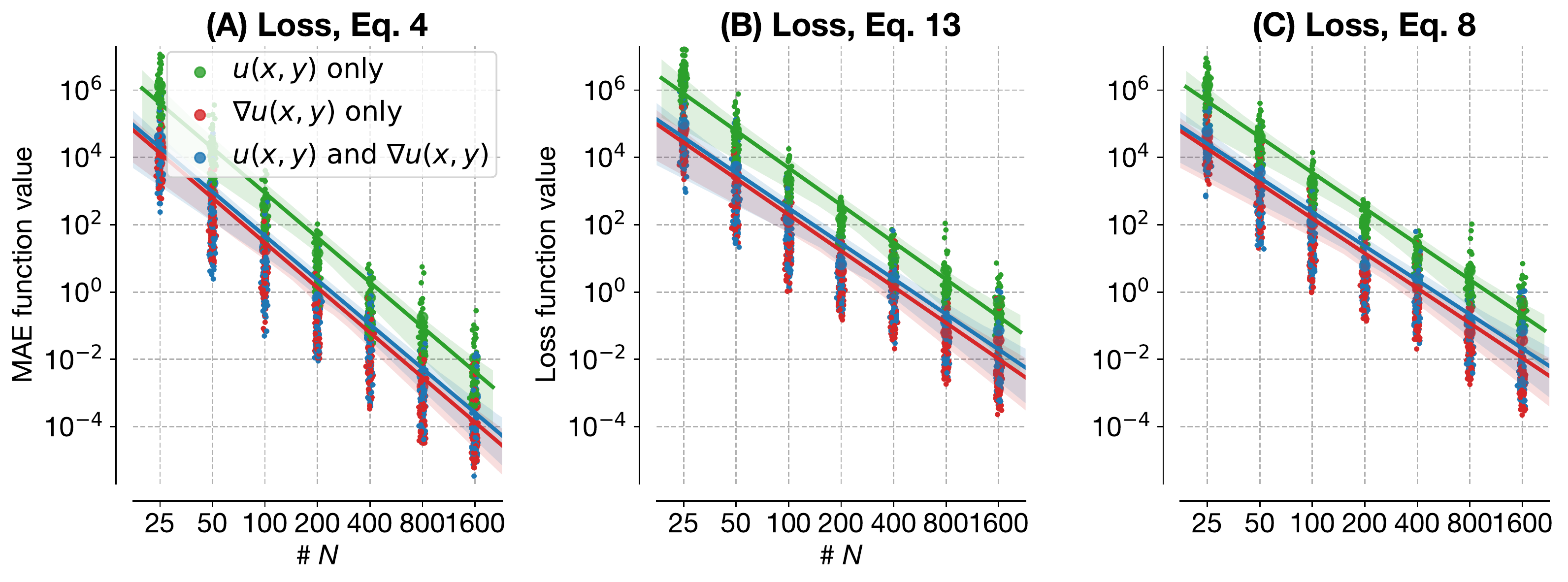}
    \caption{Learning curves for test data on three loss functions in Eq. 4 (function values only), Eq. 13 (function gradients only), and Eq. 8 (function values and gradients). Learning curves are given for three types of machines, trained on only function (green), function gradients only (red), and both function values and gradients simultaneously (blue).\label{fig:lc_fit_cost}}
\end{figure}

\begin{table}
\scriptsize
          \centering
         \caption{\label{tab:md17si}\footnotesize MAE errors for force and energy for the 10 molecules of the revised MD17 dataset. Three machines are trained for each moleucle, differening only in the training data.  "E+F" denotes machines trained on both energy and force data, "F only" denotes training on only force data, while "E only" denotes training on only energy data. Energy and force errors are given in units of kcal/mol, and kcal/mol/\AA, respectively.}
      \begin{tabular}{lllrrrrrrrr}
Molecule        & Data       & Property & 50 & 100 & 150 & 200 & 400 & 600 & 800 & 1000 \\  
\hline
Aspirin         &  E+F       & Energy  &   0.754  &  0.483  &  0.388  &  0.349  &  0.243  &  0.183  &  0.157  &  0.144 \\
                &  F only    & Energy  &   1.044  &  0.588  &  0.428  &  0.381  &  0.264  &  0.193  &  0.165  &  0.151 \\
                &  E only    & Energy  &   5.291  &  3.893  &  2.882  &  2.397  &  1.517  &  1.114  &  0.957  &  0.847 \\
                &  E+F       & Force   &   1.818  &  1.286  &  1.081  &  0.964  &  0.732  &  0.600  &  0.528  &  0.481 \\
                &  F only    & Force   &   1.824  &  1.288  &  1.082  &  0.965  &  0.738  &  0.604  &  0.528  &  0.482 \\
                &  E only    & Force   &  18.494  & 13.038  &  9.975  &  8.375  &  5.201  &  4.009  &  3.386  &  3.042 \\
\hline
Azobenzene      &  E+F       & Energy  &   0.295  &  0.196  &  0.168  &  0.153  &  0.109  &  0.084  &  0.076  &  0.064 \\
                &  F only    & Energy  &   0.302  &  0.200  &  0.169  &  0.155  &  0.110  &  0.085  &  0.076  &  0.064 \\
                &  E only    & Energy  &   2.775  &  1.580  &  1.088  &  1.009  &  0.601  &  0.476  &  0.409  &  0.368 \\
                &  E+F       & Force   &   0.844  &  0.642  &  0.551  &  0.497  &  0.379  &  0.316  &  0.280  &  0.248 \\
                &  F only    & Force   &   0.844  &  0.643  &  0.552  &  0.497  &  0.379  &  0.317  &  0.280  &  0.248 \\
                &  E only    & Force   &   9.476  &  5.381  &  3.578  &  3.235  &  2.112  &  1.743  &  1.518  &  1.369 \\
\hline
Benzene         &  E+F       & Energy  &   0.015  &  0.012  &  0.011  &  0.010  &  0.009  &  0.008  &  0.008  &  0.008 \\
                &  F only    & Energy  &   0.015  &  0.012  &  0.011  &  0.010  &  0.009  &  0.008  &  0.008  &  0.007 \\
                &  E only    & Energy  &   0.139  &  0.067  &  0.050  &  0.039  &  0.025  &  0.019  &  0.017  &  0.016 \\
                &  E+F       & Force   &   0.125  &  0.096  &  0.086  &  0.081  &  0.072  &  0.067  &  0.063  &  0.060 \\
                &  F only    & Force   &   0.125  &  0.096  &  0.086  &  0.081  &  0.072  &  0.067  &  0.063  &  0.060 \\
                &  E only    & Force   &   1.209  &  0.592  &  0.463  &  0.364  &  0.243  &  0.190  &  0.168  &  0.153 \\
\hline
Ethanol         &  E+F       & Energy  &   0.232  &  0.137  &  0.098  &  0.083  &  0.043  &  0.032  &  0.023  &  0.021 \\
                &  F only    & Energy  &   0.248  &  0.142  &  0.101  &  0.084  &  0.043  &  0.032  &  0.024  &  0.021 \\
                &  E only    & Energy  &   1.705  &  0.981  &  0.687  &  0.558  &  0.342  &  0.253  &  0.209  &  0.179 \\
                &  E+F       & Force   &   1.072  &  0.701  &  0.533  &  0.452  &  0.271  &  0.207  &  0.163  &  0.144 \\
                &  F only    & Force   &   1.088  &  0.703  &  0.534  &  0.452  &  0.272  &  0.207  &  0.166  &  0.144 \\
                &  E only    & Force   &   7.547  &  4.855  &  3.346  &  2.815  &  1.875  &  1.420  &  1.211  &  1.049 \\
\hline
Malonaldehyde   &  E+F       & Energy  &   0.312  &  0.191  &  0.136  &  0.115  &  0.070  &  0.053  &  0.042  &  0.035 \\
                &  F only    & Energy  &   0.337  &  0.213  &  0.146  &  0.122  &  0.072  &  0.054  &  0.043  &  0.035 \\
                &  E only    & Energy  &   2.290  &  1.443  &  0.954  &  0.752  &  0.460  &  0.341  &  0.281  &  0.245 \\
                &  E+F       & Force   &   1.469  &  0.966  &  0.745  &  0.629  &  0.422  &  0.330  &  0.274  &  0.237 \\
                &  F only    & Force   &   1.480  &  0.961  &  0.753  &  0.631  &  0.423  &  0.330  &  0.276  &  0.235 \\
                &  E only    & Force   &  11.690  &  6.481  &  4.851  &  3.856  &  2.488  &  1.936  &  1.629  &  1.447 \\
\hline
Naphthalene     &  E+F       & Energy  &   0.127  &  0.089  &  0.075  &  0.065  &  0.045  &  0.036  &  0.031  &  0.028 \\
                &  F only    & Energy  &   0.126  &  0.089  &  0.075  &  0.065  &  0.044  &  0.036  &  0.031  &  0.028 \\
                &  E only    & Energy  &   1.051  &  0.645  &  0.486  &  0.431  &  0.302  &  0.239  &  0.190  &  0.171 \\
                &  E+F       & Force   &   0.583  &  0.431  &  0.365  &  0.319  &  0.231  &  0.191  &  0.167  &  0.150 \\
                &  F only    & Force   &   0.583  &  0.431  &  0.366  &  0.319  &  0.231  &  0.191  &  0.168  &  0.152 \\
                &  E only    & Force   &   4.474  &  2.753  &  2.182  &  1.928  &  1.451  &  1.163  &  0.960  &  0.859 \\
\hline
Paracetamol     &  E+F       & Energy  &   0.391  &  0.244  &  0.197  &  0.172  &  0.113  &  0.091  &  0.079  &  0.067 \\
                &  F only    & Energy  &   0.401  &  0.248  &  0.200  &  0.175  &  0.115  &  0.092  &  0.080  &  0.068 \\
                &  E only    & Energy  &   4.620  &  3.027  &  2.059  &  1.654  &  0.984  &  0.673  &  0.549  &  0.470 \\
                &  E+F       & Force   &   1.181  &  0.827  &  0.694  &  0.617  &  0.442  &  0.365  &  0.321  &  0.282 \\
                &  F only    & Force   &   1.183  &  0.828  &  0.694  &  0.618  &  0.444  &  0.365  &  0.323  &  0.283 \\
                &  E only    & Force   &  16.417  & 10.726  &  7.768  &  6.123  &  3.709  &  2.687  &  2.228  &  1.922 \\
\hline
Salicylic Acid  &  E+F       & Energy  &   0.255  &  0.161  &  0.120  &  0.105  &  0.071  &  0.056  &  0.048  &  0.041 \\
                &  F only    & Energy  &   0.255  &  0.161  &  0.120  &  0.105  &  0.071  &  0.056  &  0.048  &  0.041 \\
                &  E only    & Energy  &   2.941  &  1.696  &  1.201  &  0.906  &  0.545  &  0.419  &  0.353  &  0.297 \\
                &  E+F       & Force   &   1.088  &  0.737  &  0.585  &  0.505  &  0.355  &  0.286  &  0.248  &  0.220 \\
                &  F only    & Force   &   1.089  &  0.738  &  0.585  &  0.505  &  0.355  &  0.287  &  0.248  &  0.220 \\
                &  E only    & Force   &  13.422  &  7.040  &  5.130  &  3.864  &  2.533  &  1.962  &  1.684  &  1.469 \\
\hline
Toluene         &  E+F       & Energy  &   0.169  &  0.115  &  0.097  &  0.084  &  0.060  &  0.049  &  0.044  &  0.039 \\
                &  F only    & Energy  &   0.169  &  0.117  &  0.097  &  0.084  &  0.060  &  0.049  &  0.043  &  0.038 \\
                &  E only    & Energy  &   1.401  &  0.868  &  0.642  &  0.532  &  0.344  &  0.272  &  0.232  &  0.201 \\
                &  E+F       & Force   &   0.754  &  0.545  &  0.457  &  0.404  &  0.303  &  0.251  &  0.225  &  0.204 \\
                &  F only    & Force   &   0.755  &  0.545  &  0.457  &  0.405  &  0.303  &  0.251  &  0.225  &  0.204 \\
                &  E only    & Force   &   5.997  &  3.731  &  2.965  &  2.591  &  1.691  &  1.395  &  1.212  &  1.043 \\
\hline
Uracil          &  E+F       & Energy  &   0.147  &  0.087  &  0.066  &  0.053  &  0.029  &  0.020  &  0.016  &  0.013 \\
                &  F only    & Energy  &   0.146  &  0.088  &  0.066  &  0.053  &  0.029  &  0.020  &  0.016  &  0.013 \\
                &  E only    & Energy  &   2.347  &  1.185  &  0.810  &  0.595  &  0.349  &  0.243  &  0.192  &  0.156 \\
                &  E+F       & Force   &   0.812  &  0.534  &  0.408  &  0.329  &  0.192  &  0.142  &  0.115  &  0.097 \\
                &  F only    & Force   &   0.809  &  0.535  &  0.410  &  0.329  &  0.192  &  0.142  &  0.115  &  0.097 \\
                &  E only    & Force   &  12.789  &  6.261  &  4.311  &  3.335  &  2.026  &  1.458  &  1.209  &  1.030 \\
        \hline
      \end{tabular}
\end{table}

\end{document}